\documentclass[twocolumn,aps,amsmath,prd,amssymb,eqsecnum,nofootinbib]{revtex4-1}

\usepackage{graphicx}

\begin{document}

\title{Self-force on an arbitrarily coupled static scalar particle in a wormhole space-time}

\author{Peter Taylor}
\email{ptaylor@maths.tcd.ie}
\affiliation{School of Mathematics, Trinity College, Dublin 2, Ireland}

\date{\today}
\begin{abstract}
In this paper, we consider the problem of computing the self-force and self-energy for a static scalar charge in a wormhole space-time with throat profile $r(\rho)=\sqrt{\rho^{2}+a^{2}}$ for arbitrary coupling of the field to the curvature. This calculation has previously been considered numerically by Bezerra and Khusnutdinov \cite{Khusnutdinov2}, while analytic results have been obtained in the special cases of minimal ($\xi=0$) coupling \cite{Khusnutdinov1} and conformal coupling \cite{Khusnutdinov2} ($\xi=1/8$ in three dimensions). We present here a closed form expression for the static Green's function for arbitrary coupling and hence we obtain an analytic expression for the self-force. The self-force depends crucially on the coupling of the field to the curvature of the space-time and hence it is useful to determine the dependence explicitly. The numerical computation can identify some qualitative aspects of this dependence such as the change in the sign of the force as it passes through the conformally coupled value, as well as the fact that the self-force diverges for $\xi=1/2$. From the closed form expression, it is straight-forward to see that there is an infinite set of values of the coupling constant for which the self-force diverges, but we also see that there is an infinite set of values for which the self-force vanishes. \end{abstract}
\maketitle

\section{Introduction}
Wormholes are topological bridges connecting different universes or distant regions of the same universe. Interest in wormhole space-times dates back to 1916 \cite{Flamm}, pre-dating interest in black hole space-times. Its modern popularity is owed primarily to the work of Morris and Thorne \cite{MorrisThorne} who investigated the idea of using so-called ``traversable wormholes'' as a means for time travel. Morris and Thorne showed that such space-times require a stress-energy tensor that violates the null energy condition, that is they require the existence of \textit{exotic} matter. Wormholes have subsequently played a central role in the investigation of causality violation and the status of energy conditions as  physical laws of nature. Comprehensive reviews of wormhole physics can be found in Refs.\cite{VisserWormhole, Lobo}.

We will consider a particularly simple, ultra-static, spherically symmetric wormhole geometry described by the metric
\begin{equation}
\label{eq:wormholemetric}
ds^{2}=-dt^{2}+d\rho^{2}+r^{2}(\rho)d\Omega^{2},
\end{equation}
where $d\Omega^{2}=d\theta^{2}+\sin^{2}\theta \,d\phi^{2}$ is the line-element on the two-sphere $\mathbb{S}^{2}$ and $r(\rho)=\sqrt{\rho^{2}+a^{2}}$ is the profile of the wormhole throat with a minimum radius $r(0)=a$. The range of the radial coordinate is the entire real line, $-\infty<\rho<\infty$, and the throat connects two identical asymptotically flat space-times. The manifold is everywhere smooth with scalar curvature given by
\begin{equation}
\label{eq:curvature}
R= -\frac{2 a^{2}}{(\rho^{2}+a^{2})^{2}}.
\end{equation}

We consider the problem of computing the self-force on a scalar charge at rest in the metric described by (\ref{eq:wormholemetric}), allowing for the arbitrary values of the coupling constant in the wave equation. A similar calculation was considered by Khusnutdinov and Bakhmatov \cite{Khusnutdinov1} who computed the electrostatic self-force on a charged particle at rest this wormhole space-time. The electrostatic wave equation is equivalent to that of a minimally coupled scalar charge at rest and hence the self-force on a static scalar charge is equal to the electrostatic sef-force, up to an overall sign. Bezerra and Khusnutdinov \cite{Khusnutdinov2} numerically evaluated the self-force on a static scalar for arbitrary coupling, though we disagree with the overall sign of their results. The electrostatic case was reconsidered by Linet \cite{LinetWormhole1} who derived the Green's function in closed form by transforming to isotropic coordinates and expanding about the Euclidean distance in these coordinates, a method first adopted by Copson \cite{Copson} in deriving the electrostatic potential in the Schwarzschild space-time.

The self-force is obtained by taking the gradient of the retarded field which is singular at the particle's location and requires regularization. In order to regularize the self-force, we compute the Detweiler-Whiting singular field, which upon subtraction yields a quantity that is regular at the scalar charge's location. We obtain an analytic expression for the self-force for arbitrary values of the coupling constant, which reveals some expected features such as infinite poles which also occurs in the expression for the self-force \cite{Khusnutdinov2} for the wormhole with throat profile $r(\rho)=|\rho|+a$, but we also find some unexpected features such as an infinite set of values for which the self-force vanishes.

\section{Green's Function}
The Green's function satisfies the following inhomogeneous wave equation,
\begin{equation}
(\Box-\xi \, R)G(x,x')=-g^{-1/2}\delta(x-x'),
\end{equation}
where $\Box$ is the d'Alembertian wave operator on the metric (\ref{eq:wormholemetric}), and $\xi$ is the coupling to the scalar curvature which is given by Eq.(\ref{eq:curvature}). For the charge at rest according to observers moving on integral curves of the Killing vector $\partial/\partial t$, the wave equation reduces to the three-dimensional Helmholtz equation,
\begin{align}
\Big\{\frac{\partial}{\partial \rho}\Big((\rho^{2}+a^{2})\frac{\partial}{\partial \rho}\Big)+\frac{1}{\sin\theta}\frac{\partial}{\partial\theta}\Big(\sin\theta \frac{\partial}{\partial\theta}\Big)+\frac{1}{\sin^{2}\theta}\frac{\partial^{2}}{\partial\phi^{2}}\nonumber\\
+\frac{2\xi a^{2}}{(\rho^{2}+a^{2})}\Big\}G^{(3)}(\textbf{x},\textbf{x}')=-\frac{\delta(\textbf{x}-\textbf{x}')}{\sin\theta}.
\end{align}
The Green's function may be given as a mode-sum over separable solutions to the homogeneous equation,
\begin{equation}
G^{(3)}(\textbf{x}, \textbf{x}')=\frac{1}{4\pi}\sum_{l=0}^{\infty}(2l+1)P_{l}(\cos\gamma)g_{l}(\rho, \rho'),
\end{equation}
where $P_{l}(x)$ is the Legendre polynomial, $\cos\gamma=\cos\theta\cos\theta'+\sin\theta\sin\theta'\cos\Delta\phi$ and $g_{l}(\rho,\rho')$ satisfies the inhomogeneous radial equation,
\begin{align}
\Big\{\frac{d}{d\rho}\Big((\rho^{2}+a^{2})\frac{d}{d\rho}\Big)-l(l+1)+\frac{2\xi a^{2}}{(\rho^{2}+a^{2})}\Big\}g_{l}(\rho,\rho')\nonumber\\
=-\delta(\rho-\rho').
\end{align}
With the transformation
\begin{equation}
y=\rho/a,
\end{equation}
the radial equation takes a more simple form,
\begin{align}
\label{eq:radialeq}
\Big\{\frac{d}{d y}\Big((y^{2}+1)\frac{d}{d y}\Big)-l(l+1)+\frac{2\xi}{(y^{2}+1)}\Big\}g_{l}(y,y')\nonumber\\
=-\frac{1}{|a|}\delta(y-y').
\end{align}
The general solution may be written as a normalized product of two linearly independent solutions of the homogeneous equation
\begin{equation}
g_{l}(y,y')=\frac{1}{|a| }\frac{\Psi_{l}^{(1)}(y_{<})\Psi_{l}^{(2)}(y_{>})}{N},
\end{equation}
where $y_{<}=\min\{y,y'\}$, $y_{>}=\max\{y,y'\}$ and the normalization constant $N$ is determined by the Wronskian of the two solutions. The boundary conditions on the Green's function at $\rho\rightarrow\pm\infty$ require
\begin{align}
\Psi^{(1)}_{l}(y)&\rightarrow 0,\quad\textrm{as}\,\,y\rightarrow-\infty,\nonumber\\
\Psi^{(2)}_{l}(y)&\rightarrow 0,\quad\textrm{as}\,\,y\rightarrow \infty.
\end{align}

Writing $z=i\,y$, it is clear that the solutions of the homogeneous equation (\ref{eq:radialeq}) are associated Legendre functions of pure imaginary order,
\begin{equation}
P_{l}^{\pm\mu}(\pm i y), \, Q_{l}^{\pm\mu}(\pm i y),\qquad\textrm{where}\,\,\mu=\sqrt{2\xi}.
\end{equation}
The multi-valuedness of the associated Legendre functions gives rise to a discontinuity at $y=0$. For $y\ge 0$, we choose $\Psi^{(2)}_{l}(y)=Q^{\mu}_{l}(i y)$ which vanishes as $y\rightarrow \infty$ as required. However, for the solution to be continuous across $y=0$, we take the branch obtained from the principal branch by encircling the branch point $z=1$ (but not the point $z=-1$) once. If we denote this branch by $Q^{\mu}_{l,1}(z)$, then it can be shown \cite{Olver}
\begin{align}
Q^{\mu}_{l,1}(z)=e^{-\mu\pi i} Q^{\mu}_{l}(z)-i\pi\,e^{\mu\pi i}\frac{\Gamma(l+\mu+1)}{\Gamma(l-\mu+1)}P^{-\mu}_{l}(z).
\end{align}
Then, the function
\begin{align}
\Psi^{(2)}_{l}(y)=\begin{cases}
Q^{\mu}_{l}(i y)\quad \quad\qquad\qquad\qquad\qquad\qquad\quad y\ge 0,\nonumber\\\nonumber\\
\displaystyle{e^{-\mu\pi i} Q^{\mu}_{l}(i y)-i\pi\,e^{\mu\pi i}\frac{\Gamma(l+\mu+1)}{\Gamma(l-\mu+1)}P^{-\mu}_{l}(i y)}\nonumber\\
 \qquad\qquad\qquad\qquad\qquad\qquad\qquad\qquad y<0,
\end{cases}
\end{align}
is continuous for $-\infty<y<\infty$ and satisfies the appropriate boundary condition and hence is the correct choice.

The symmetry of the space-time implies that we take our inner solution to be $\Psi^{(1)}_{l}(y)=\Psi^{(2)}_{l}(-y)$, or given explicitly,
\begin{equation}
\Psi^{(1)}_{l}(y)=\begin{cases}
\displaystyle{e^{-\mu\pi i} Q^{\mu}_{l}(-i y)-i\pi\,e^{\mu\pi i}\frac{\Gamma(l+\mu+1)}{\Gamma(l-\mu+1)}P^{-\mu}_{l}(-i y)}\nonumber\\
 \qquad\qquad\qquad\qquad\qquad\qquad\qquad\qquad\,\,\,\, y\ge0,\nonumber\\\nonumber\\
 Q^{\mu}_{l}(-i y)\quad \quad\qquad\qquad\qquad\qquad\qquad\quad y< 0.
\end{cases}
\end{equation}
This solution clearly satisfies the vanishing boundary condition at $y\rightarrow-\infty$ ($\rho\rightarrow-\infty$) and is linearly independent to $\Psi_{l}^{(1)}(y)$ in the two regions of the space-time. We can re-write the solution in a more symmetric form using standard results for the Legendre functions \cite{gradriz}, yielding
\begin{align}
\Psi^{(1)}_{l}(y)=\begin{cases}
\displaystyle{(-1)^{l+1} e^{\mu\pi i}\Big[ Q^{\mu}_{l}(i y)+i\pi\,\frac{\Gamma(l+\mu+1)}{\Gamma(l-\mu+1)}P^{-\mu}_{l}(i y)\Big]}\nonumber\\
 \qquad\qquad\qquad\qquad\qquad\qquad\qquad\qquad\,\,\, \,y\ge0,\nonumber\\\nonumber\\
 (-1)^{l+1}Q^{\mu}_{l}(i y)\quad \quad\qquad\qquad\qquad\qquad\,\, y< 0.
\end{cases}
\end{align}

The normalization constant is given by
\begin{align}
N&=-(y^{2}+1)W\{\Psi^{(1)}_{l}(y), \Psi^{(2)}_{l}(y)\}\nonumber\\
&=\pi (-1)^{l+1}e^{2\mu \pi i}\frac{\Gamma(l+\mu+1)}{\Gamma(l-\mu+1)}.
\end{align}
So the radial solution may be written as
\begin{align}
&g_{l}(y,y')=\nonumber\\
&\begin{cases}
\displaystyle{\frac{e^{\mu\pi i}}{|a|\,\pi}Q^{-\mu}_{l}(iy)Q^{\mu}_{l}(iy')}+\frac{i \,e^{-\mu\pi i}}{|a|}P^{-\mu}_{l}(iy_{<})Q^{\mu}_{l}(iy_{>}) \nonumber\\
\quad\quad\quad\qquad\qquad\qquad\qquad\qquad \textrm{if }\,y,y'\ge0,\nonumber\\\nonumber\\
\displaystyle{\frac{e^{-\mu\pi i}}{|a| \pi}Q^{-\mu}_{l}(iy)Q^{\mu}_{l}(iy')}-\frac{i \,e^{-\mu\pi i}}{|a|}Q^{\mu}_{l}(iy_{<})P^{-\mu}_{l}(iy_{>}) \nonumber\\
\quad\quad\quad\qquad\qquad\qquad\qquad\qquad \textrm{if }\,y,y'<0,\nonumber\\\nonumber\\
\displaystyle{\frac{1}{|a|\pi}Q^{-\mu}_{l}(iy)Q^{\mu}_{l}(iy')} \qquad\quad\quad\,\,\textrm{if }\, y\ge0, y'<0\nonumber\\
\quad\quad\quad\qquad\qquad\qquad\qquad\qquad \textrm{or }y'\ge 0, y<0.
\end{cases}
\end{align}

Let us assume, without loss of generality, that $y, y'\ge 0$ (the closed-form expression won't depend on this condition), then the Green's function is
\begin{align}
\label{eq:gmodesum}
&G^{(3)}(\textbf{x},\textbf{x}')=\frac{i}{4\pi |a|}\sum_{l=0}^{\infty}(2l+1)P_{l}(\cos\gamma)\nonumber\\
&\times\Big[e^{-\mu\pi i}P^{-\mu}_{l}(iy_{<})Q^{\mu}_{l}(iy_{>})+\frac{1}{\pi i}e^{\mu\pi i}Q^{-\mu}_{l}(iy)Q^{\mu}_{l}(iy')\Big].
\end{align}
In a recent paper \cite{OttewillTaylor4}, we have derived the following summation formula for associated Legendre functions of arbitrary complex order,
\begin{align}
\label{eq:legendresum}
\sum_{l=0}^{\infty}(2l+1)P_{l}(\cos\gamma)e^{-i\mu\pi}P^{-\mu}_{l}(\eta_{<})Q^{\mu}_{l}(\eta_{>})=\frac{e^{-\mu\cosh^{-1}(\chi)}}{R^{1/2}}
\end{align}
where
\begin{align}
R&=\eta^{2}+\eta'^{2}-2\eta\eta'\cos\gamma-\sin^{2}\gamma,\nonumber\\
\chi&=\frac{\eta\eta'-\cos\gamma}{(\eta^{2}-1)^{1/2}(\eta'^{2}-1)^{1/2}},
\end{align}
and where the radial variable here is real and runs over the range $\eta>1$. We can analytically continue this result by making the transformation $\eta=i y$ and taking the appropriate branch. We obtain
\begin{align}
\label{eq:legendresumcomplex}
&\sum_{l=0}^{\infty}(2l+1)P_{l}(\cos\gamma)e^{-\mu\pi i}P^{-\mu}_{l}(iy_{<})Q^{\mu}_{l}(iy_{>})\nonumber\\
&=\begin{cases}
\displaystyle{\frac{e^{\mu\cosh^{-1}(\chi)}}{i R^{1/2}}}\quad &y,y'\ge 0,\nonumber\\\nonumber\\
\displaystyle{-\frac{e^{-\mu\cosh^{-1}(\chi)}}{i R^{1/2}}}\quad &y,y'<0, \nonumber\\\nonumber\\
\displaystyle{\frac{e^{-\mu\pi i}e^{\mu\cosh^{-1}(\chi)}}{i R^{1/2}}}\quad &y>0, y'<0\,\,\textrm{or}\,\,y<0, y'>0,
\end{cases}
\end{align}
where
\begin{align}
R&=y^{2}+y'^{2}-2y y'\cos\gamma+\sin^{2}\gamma,\nonumber\\
\chi&=\frac{y y'+\cos\gamma}{(y^{2}+1)^{1/2}(y'^{2}+1)^{1/2}}.
\end{align}
These results may be checked numerically by multiplying both sides by $P_{l'}(\cos\gamma)$ and integrating with respect to $\gamma$.

For $y, y'\ge0$, we can use the well-known relations between Legendre functions to rewrite the Green's function (\ref{eq:gmodesum}) as
\begin{align}
&G^{(3)}(\textbf{x},\textbf{x}')=\frac{1}{8\pi |a|\sin(\mu \pi)}\sum_{l=0}^{\infty}(2l+1)P_{l}(\cos\gamma)\nonumber\\
&\times \Big[ -e^{-2\mu\pi i}P^{-\mu}_{l}(i y_{<}) Q^{\mu}_{l}(i y_{>})+e^{2\mu \pi i} P^{\mu}_{l}(i y_{<})Q^{-\mu}_{l}( i y_{>})\Big].
\end{align}
We can employ the summation formula (\ref{eq:legendresumcomplex}) to obtain the following closed form representation of the Green's function for a static scalar charge in our wormhole space-time
\begin{align}
\label{eq:gclosed}
G^{(3)}(\textbf{x},\textbf{x}')=\frac{1}{4\pi |a|\sin(\mu \pi)}\frac{\sin(\mu\cos^{-1}(-\chi))}{R^{1/2}},
\end{align}
where we have used the fact that
\begin{equation}
\pi+i\,\cosh^{-1}(\chi)=\cos^{-1}(-\chi)\quad\textrm{for}\quad |\chi|<1.
\end{equation}
A similar analysis for the case where $y, y'<0$ and where the two points are in different regions of the space-time yield the same closed form expression as Eq.(\ref{eq:gclosed}). Finally, restoring the variable $\rho= y/a$ gives
\begin{align}
\label{eq:gclosednew}
&G^{(3)}(\textbf{x},\textbf{x}')\nonumber\\
&=\frac{1}{4\pi \sin(\mu \pi)}\frac{\sin(\mu\cos^{-1}(-\chi))}{(\rho^{2}+\rho'^{2}-2\rho\rho'\cos\gamma+a^{2}\sin^{2}\gamma)^{1/2}},
\end{align}
where
\begin{equation}
\chi=\frac{\rho\rho'+a^{2} \cos\gamma}{(\rho^{2}+a^{2})^{1/2}(\rho'^{2}+a^{2})^{1/2}}.
\end{equation}

We note that this Green's function has the correct Hadamard singularity structure \cite{Hadamard} in three dimensions. We also note that the Green's function possesses infinite poles for certain values of the coupling constant given by
\begin{equation}
\label{eq:xidiv}
\xi=\frac{n^{2}}{2},\quad n\in \mathbb{Z}\setminus\{0\}.
\end{equation}
This is analogous to the infinite poles that arise in the Green's function for throat profile $r(\rho)=|\rho|+a$, as shown in Ref.~\cite{Khusnutdinov2}.

For minimal coupling, Eq.(\ref{eq:gclosednew}) is understood to mean
\begin{equation}
G^{(3)}_{\mu=0}=\frac{1}{4\pi^{2}}\frac{\cos^{-1}(-\chi)}{(\rho^{2}+\rho'^{2}-2\rho\rho'\cos\gamma+a^{2}\sin^{2}\gamma)^{1/2}},
\end{equation}
which agrees with the closed form representation given in Ref.~\cite{Khusnutdinov1} if we take $\gamma=0$.

\section{Self force and self energy}

There are a number of approaches one can take to compute the self-force (see \cite{PoissonLR} for example), but the most direct way for a static charge is by subtracting the Detweiler-Whiting \cite{DetweilerWhiting2003} parametrix which yields a finite quantity upon taking coincidence limits. It is constructed in such a way as to leave, upon subtraction from the Green's function, only the radiative part of the field entirely responsible for the self-force. This approach was adopted in Ref.~\cite{OttewillTaylor3} to compute the self-force on a static scalar charge in Kerr space-time. There it was shown that for a static scalar charge with world-line $x'^{\alpha}=z^{\alpha}(\tau)$ and four-velocity $u^{\alpha}=u^{t'}\delta^{\alpha}{}_{t'}$ in a general stationary space-time, the self-force is
\begin{align}
\label{eq:fselfstatic}
f^{\textrm{self}}_{\alpha}=4\pi q^{2}\lim_{\textbf{x}\rightarrow \textbf{x}'}\Big[\nabla_{\alpha}\Big(\frac{1}{u^{t'}}G^{(3)}(\textbf{x}, \textbf{x}')-G_{\textrm{DW}}(\textbf{x}, \textbf{x}')\Big)\Big]
\end{align}
where $G^{(3)}(\textbf{x},\textbf{x}')$ is the zero-frequency mode of the four-dimensional retarded Green's function (modulo a factor of $2\pi$) and $G_{\textrm{DW}}(x, x')$ is the Detweiler-Whiting Green's function given by \cite{PoissonLR}
\begin{align}
\label{eq:gdw}
G_{\textrm{DW}}(x, x')&=\frac{1}{4\pi}\Big(\frac{\Delta^{1/2}(x, x_{\textrm{ret}})}{2\,r_{\textrm{ret}}}+\frac{\Delta^{1/2}(x, x_{\textrm{adv}})}{2\,r_{\textrm{adv}}}\nonumber\\
&+\frac{1}{2}\int_{\tau_{\textrm{ret}}}^{\tau_{\textrm{adv}}}V(x, x'(\tau))d\tau\Big),
\end{align}
where $\Delta$ is the Van-Vleck Morrette determinant, $r_{\textrm{ret}}$ is the retarded distance between the field point $x$ and the retarded point $x'=x_{\textrm{ret}}$ on the charge's world-line and $r_{\textrm{adv}}$ is the advanced distance between $x$ and the advanced point $x'=x_{\textrm{adv}}$ on the world-line.

In the case of minimal coupling, it has recently been shown \cite{CasalsPoissonVega} that the Detweiler-Whiting Green's function for a static charge in a static space-time is equivalent to the direct part of the Hadamard Green's function on the spatial part of the metric up to the order required for regularization, and equal up to all orders for ultra-static space-times such as the wormhole space-time under consideration. In that case, the appropriate parametrix that must be subtracted to give the correct self-force is particularly simple,
\begin{equation}
\label{eq:gdwmin}
G_{\textrm{DW}}(\textbf{x},\textbf{x}')=\frac{1}{4\pi}\frac{\Delta^{1/2}(\textbf{x},\textbf{x}')}{\sqrt{2\sigma(\textbf{x},\textbf{x}')}}
\end{equation}
where $\Delta$ and the world-function $\sigma$ are calculated on the three-dimensional metric
\begin{equation}
ds_{(3)}^{2}=d\rho^{2}+r^{2}(\rho)d\Omega^{2}.
\end{equation}
However, it is clear that this statement cannot be true for arbitrary coupling since the biscalar $V(x,x')$ appearing in Eq.~(\ref{eq:gdw}) depends on the coupling constant $\xi$ while the direct part of the three-dimensional Hadamard form (\ref{eq:gdwmin}) does not, indeed it is purely geometrical. 

\begin{figure*}
\centering
\includegraphics[width=10cm]{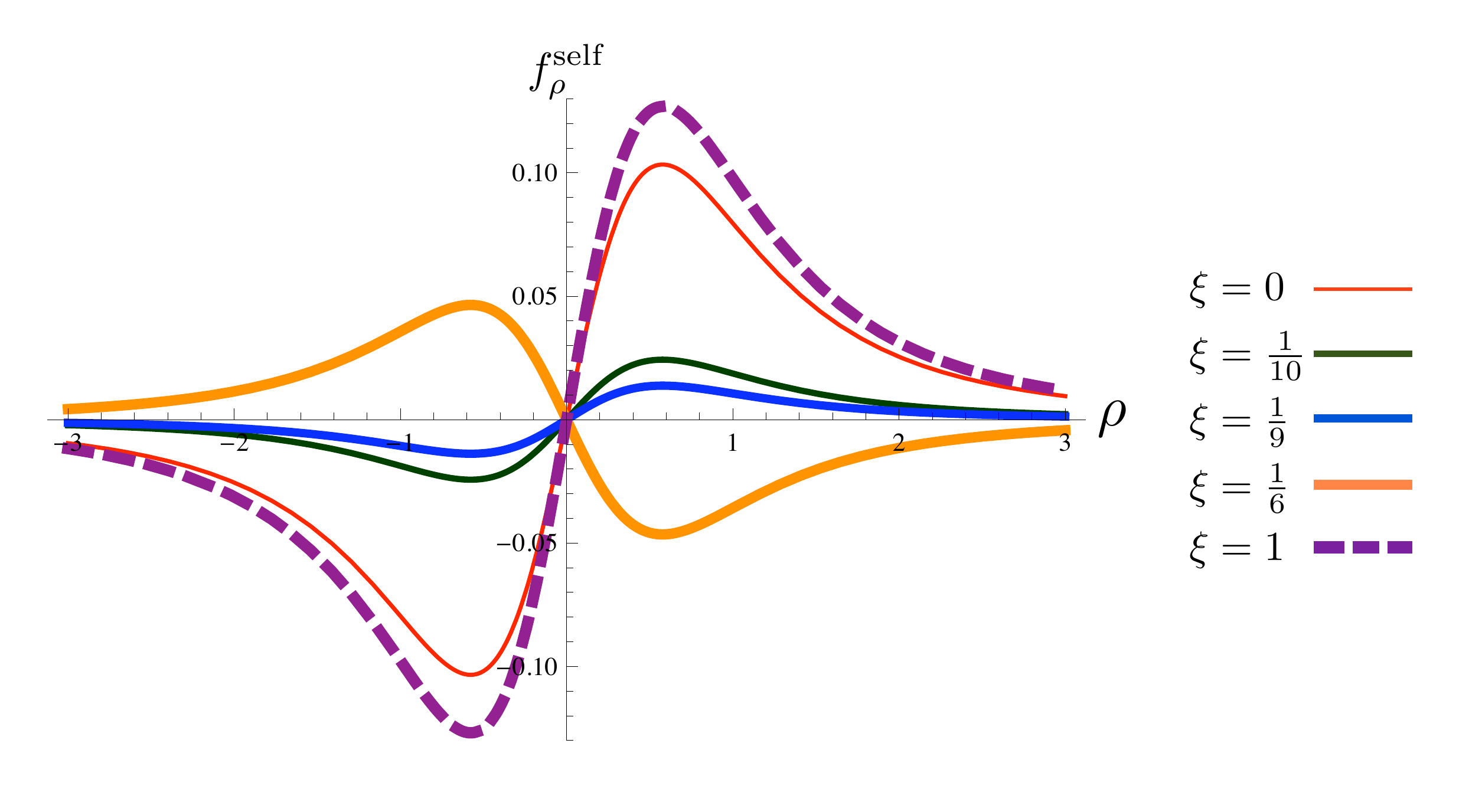}
\caption{{Plot of the radial component of the self-force on a static scalar charge in a wormhole space-time. We have set $q=1$ and $a=1$. The thickness of the curves increases as the coupling increases and the continuous lines are for values in the range $\xi<1/2$.} }
\label{fig:selfforce}
\end{figure*} 

In obtaining an expression for $G_{\textrm{DW}}(x,x')$, we will use the standard expansions
\begin{align}
\Delta^{1/2}(x,x')&=1+\tfrac{1}{12}R_{a b}\sigma^{;a}\sigma^{;b}+\mathcal{O}(\sigma^{3/2})\nonumber\\
V(x,x')&=\tfrac{1}{2}(\xi-\tfrac{1}{6})R+\mathcal{O}(\sigma^{1/2}).
\end{align}
We can use the fact that the space-time is spherically symmetric to set $\gamma=0$, in which case the Van-Vleck Morrette expansion reduces to
\begin{equation}
\Delta^{1/2}(\rho,\rho')=1-\frac{a^{2}}{6(\rho'^{2}+a^{2})^{2}}\Delta\rho^{2}+\mathcal{O}(\Delta\rho^{3}).
\end{equation}
From the expansion for $V(x,x')$, the integral in Eq.(\ref{eq:gdw}) is
\begin{equation}
\frac{1}{2}\int_{\tau_{\textrm{ret}}}^{\tau_{\textrm{adv}}}V(x, x'(\tau))d\tau=\tfrac{1}{4}(\xi-\tfrac{1}{6}) R\,\Delta\tau+\mathcal{O}(\Delta x^{2}),
\end{equation}
where $\Delta\tau=\tau_{\textrm{adv}}-\tau_{\textrm{ret}}$. Coordinate expansions for the quantities $r_{\textrm{ret}}$, $r_{\textrm{adv}}$ and $\Delta\tau$ were computed for a static charge in an arbitrary stationary space-time in Ref.~\cite{OttewillTaylor3}, which in the wormhole space-time under consideration, simplify greatly to give
\begin{equation}
r_{\textrm{ret}}=r_{\textrm{adv}}=\tfrac{1}{2}\Delta\tau=|\Delta\rho|.
\end{equation}
Hence the Detweiler-Whiting parametrix is
\begin{equation}
\label{eq:gdwexp}
G_{\textrm{DW}}(\rho,\rho')=\frac{1}{4\pi\,|\Delta\rho|}-\frac{\mu^{2}a^{2}}{(\rho'^{2}+a^{2})^{2}}|\Delta\rho|+\mathcal{O}(\Delta\rho^{2}),
\end{equation}
where recall that $\mu=\sqrt{2\xi}$ and we have used the expression for the scalar curvature given by Eq.(\ref{eq:curvature}).

We also require an expansion for the closed form Green's function of Eq.(\ref{eq:gclosednew}). Taking the scalar particle's location to be $\rho'$ and the field point to be $\rho$, then for $\rho$ near $\rho'$ the Green's function may be expanded as
\begin{align}
\label{eq:gexp}
G^{(3)}(\rho,\rho')&=\frac{1}{4\pi \Delta\rho}-\frac{1}{4\pi}\frac{\mu\cos(\mu\pi)}{\sin(\mu\pi)}\frac{|a|}{(\rho'^{2}+a^{2})}\nonumber\\
&-\frac{1}{8\pi}\frac{|a|\mu}{\sin(\mu\pi)}\Big(\frac{2\rho'\cos(\mu\pi)+|a|\mu \sin(\mu\pi)}{(\rho'^{2}+a^{2})^{2}}\Big)\Delta\rho\nonumber\\
&+\mathcal{O}(\Delta\rho^{2}),
\end{align}
where $\Delta\rho=\rho'-\rho$ which we assume, without loss of generality, to be positive.

The self-energy for a charge $q$ is proportional to the coincidence limit of the radiative field, and is given by
\begin{align}
U(\rho')&=2\pi q^{2}\lim_{\rho\rightarrow\rho'}\Big[G^{(3)}(\rho,\rho')-G_{\textrm{DW}}(\rho,\rho')\Big]\nonumber\\
&=-\frac{q^{2}}{2}\frac{\mu\cos(\mu\pi)}{\sin(\mu\pi)}\frac{|a|}{(\rho'^{2}+a^{2})}.
\end{align}
For minimal coupling, $\xi=0$ ($\mu=0$), we arrive at the result of Khusnutdinov and Bahkmatov \cite{Khusnutdinov1} (which was rederived by Linet \cite{LinetWormhole1}),
\begin{equation}
U(\rho')=-\frac{q^{2}|a|}{2\pi (\rho'^{2}+a^{2})},
\end{equation}
while for conformal coupling $\xi=1/8$ ($\mu=1/2$), the self-energy vanishes, which agrees with the results of Bezerra and Khusnutdinov \cite{Khusnutdinov2}. The self-energy, like the Green's function, is also divergent for the particular values of the coupling constant given by Eq.~(\ref{eq:xidiv}).

Turning now to the calculation of the self-force, which we have defined for a static scalar charge in a stationary space-time in Eq.(\ref{eq:fselfstatic}).  For the ultra-static wormhole space-time under consideration, this reduces to
\begin{equation}
f^{\textrm{self}}_{\rho}=4\pi q^{2} \lim_{\rho\rightarrow\rho'}\nabla_{\rho}\Big(G^{(3)}(\rho, \rho')-G_{\textrm{DW}}(\rho, \rho')\Big),
\end{equation}
with all other components vanishing due to the spherical symmetry. Substituting in the expression for the singular field (\ref{eq:gdwexp}) and the Green's function expansion (\ref{eq:gexp}), we obtain for the self-force
\begin{equation}
f^{\textrm{self}}_{\rho}=q^{2}|a|\mu\cot(\mu\pi)\frac{\rho'}{(\rho'^{2}+a^{2})^{2}}.
\end{equation}
Again, we note that the self-force has an infinite number of poles  whenever the coupling constant is $\xi=n^{2}/2$, where $n\in \mathbb{Z}/\{0\}$. A similar analysis \cite{Khusnutdinov2} for throat profile $r(\rho)=|\rho|+a$ also exhibits this behaviour for the self-force.

For minimal coupling, we obtain
\begin{equation}
f^{\textrm{self}}_{\rho}=\frac{q^{2}\rho' |a|}{\pi(\rho'^{2}+a^{2})^{2}},
\end{equation}
which is in agreement with the electrostatic self-force derived in Ref.~\cite{Khusnutdinov1}, modulo the sign of the force since the field for a minimally coupled static scalar in an ultra-static space-time is minus the electrostatic field. Hence the self-force on a static scalar in the wormhole space-time is always repulsive with respect to the throat whereas in the electrostatic case, it is always attractive with respect to the throat. 

For the conformal coupling in three dimensions , $\xi=1/8$, we obtain zero self-force which is in agreement with the result of Ref.~\cite{Khusnutdinov2}. In fact, just as there are an infinite set of values for which the self-force diverges, there are also an infinite set for which it vanishes,
\begin{equation}
 f^{\textrm{self}}_{\rho}=0,\quad \textrm{for}\,\,\mu=\frac{2n+1}{2}\iff \xi=\frac{(2n+1)^{2}}{8}.
\end{equation}
That the self-force vanishes for $\xi=1/8$ for a massless field is expected since the spatial section of the metric is conformally flat. However, the existence of an infinite set of values of the coupling constant for which the self-force vanishes is a surprising result, at least to this author.
 
In Fig.~\ref{fig:selfforce}, we plot the self-force for various values of the coupling constant. The first thing to note is that the overall sign of the self-force differs to the numerical computation of Ref.~\cite{Khusnutdinov2}. So for the field with conformal coupling in the range $0<\xi<1/8$, the self-force is everywhere repulsive with respect to the wormhole throat, not attractive as previously claimed. The self-force vanishes for conformal coupling $\xi=1/8$ and then becomes attractive for values in the range $1/8<\xi<1/2$. The force becomes increasingly attractive with respect to the throat as $\xi$ increases towards $\xi=1/2$ where it becomes divergent. This cycle then continues, the force is repulsive for $1/2<\xi<9/8$, vanishing at $\xi=9/8$ and attractive with respect to the throat for $9/8<\xi< 2$ and diverges for $\xi=2$ etc. 

These poles in the expression for the self-force, which also appear in the closed form representation of the Green's function, can be understood in the context of quantum mechanical scattering theory as the energy of bound states. In this classical context however, the existence of such poles suggests that for certain coupling strengths of the scalar field to the curvature, the charge cannot remain static.

For a particular value of the coupling constant, the direction of the self-force in either part of the space-time is independent of the radius at which the scalar charge is being held. For  example, for $1/8<\xi<1/2$ the force is always attractive with respect to the throat regardless of where in the wormhole space-time the charge is placed, and hence for the coupling strength in this range we may hypothesize that the scalar charges will accumulate in the vicinity of the throat. On the other hand, for $\xi<1/8$, the self-force is everywhere repulsive with respect to the throat. The magnitude of the self-force for a static scalar is maximized when the charge is held at $\rho=\pm |a|/\sqrt{3}$, and the force at this radius is
\begin{equation}
f^{\textrm{self}}_{ \textrm{max}}=\pm \frac{3\sqrt{3}q^{2}\mu\cot(\mu\pi)}{16 a^{2}}.
\end{equation}

\section{Conclusions}
We have obtained an analytic expression for the self-force on an arbitrarily coupled static scalar charge in a wormhole space-time with throat profile $r(\rho)=\sqrt{\rho^{2}+a^{2}}$. Analytic expressions have previously been obtained only for minimally and conformally coupled scalar fields, while the case of general coupling had been computed numerically. We find that there are infinite poles in the expression for the self-force corresponding to the values of the coupling constant where $\xi=n^{2}/2$, for $n\in\mathbb{Z}\setminus\{0\}$. We also find that there are an infinite set of values of the coupling constant for which the self-force vanishes.

\section*{Acknowledgements}
I am grateful to Adrian Ottewill and Marc Casals for reading the manuscript and for their insightful suggestions. 

I am also indebted to Bernard Linet for his friendly correspondence and for pointing out an error in an earlier version of this manuscript.

\bibliographystyle{apsrev}
\bibliography{database}
\end{document}